\documentclass[a4paper,12pt]{article}
\pdfoutput=1
\usepackage{graphicx,rotating,amssymb,amsmath,cite}
\usepackage[normalem]{ulem}

\ifx\pdfoutput\undefined
\usepackage[dvips,bookmarks]{hyperref}	% This is for arXiv.org
\else
\usepackage{hyperref}	% This is for pdftex
\fi
\hypersetup{colorlinks,bookmarksopen,bookmarksnumbered,citecolor=rossos,
linkcolor=verde,pdfstartview=FitH,urlcolor=rossos}

\def\hhref#1{\href{http://arxiv.org/abs/#1}{#1}} % in bibliography
\usepackage{footnote}
\usepackage{booktabs}
\usepackage{multicol}
\usepackage{bold-extra}
\usepackage{color}
\definecolor{rosso}{cmyk}{0,1,1,0.4}
\definecolor{rossos}{cmyk}{0,1,1,0.55}
\definecolor{rossoc}{cmyk}{0,1,1,0.2}
\definecolor{blu}{cmyk}{1,1,0,0.3}
\definecolor{blus}{cmyk}{1,1,0,0.6}
\definecolor{bluc}{cmyk}{1,1,0,0.1}
\definecolor{verde}{cmyk}{0.92,0,0.59,0.25}
\definecolor{verdec}{cmyk}{0.92,0,0.59,0.15}
\definecolor{verdes}{cmyk}{0.92,0,0.59,0.4}

\font\tenrsfs=rsfs10 at 12pt
\font\sevenrsfs=rsfs7
\font\fiversfs=rsfs5
\newfam\rsfsfam
\textfont\rsfsfam=\tenrsfs
\scriptfont\rsfsfam=\sevenrsfs
\scriptscriptfont\rsfsfam=\fiversfs
\def\mathscr#1{{\fam\rsfsfam\relax#1}}

\def\circa#1{\,\raise.3ex\hbox{$#1$\kern-.75em\lower1ex\hbox{$\sim$}}\,}

\newcommand{\beq}{\begin{equation}}
\newcommand{\eeq}{\end{equation}}

\def\circa#1{\,\raise.3ex\hbox{$#1$\kern-.75em\lower1ex\hbox{$\sim$}}\,}
\makeatletter

\def\art{\@ifnextchar[{\eart}{\oart}}
\def\eart[#1]#2#3#4#5#6{{\rm #2}, {#3 #4} {\rm (#6) #5} [{\hhref{#1}}]}
\def\hepart[#1]#2{{\rm #2, \hhref{#1}}}
\newcommand{\oart}[5]{{\rm #1}, {#2 #3} {\rm (#5) #4}}

\newcounter{alphaequation}[equation]
\def\thealphaequation{\theequation\hbox to
0.6em{\hfil\alph{alphaequation}\hfil}}
\def\eqnsystem#1{
\def\@eqnnum{{\rm (\thealphaequation)}}
\def\@@eqncr{\let\@tempa\relax \ifcase\@eqcnt \def\@tempa{& & &} \or
  \def\@tempa{& &}\or \def\@tempa{&}\fi\@tempa
  \if@eqnsw\@eqnnum\refstepcounter{alphaequation}\fi
\global\@eqnswtrue\global\@eqcnt=0\cr}
\refstepcounter{equation} \let\@currentlabel\theequation \def\@tempb{#1}
\ifx\@tempb\empty\else\label{#1}\fi
\refstepcounter{alphaequation}
\let\@currentlabel\thealphaequation
\global\@eqnswtrue\global\@eqcnt=0 \tabskip\@centering\let\\=\@eqncr
$$\halign to \displaywidth\bgroup \@eqnsel\hskip\@centering
$\displaystyle\tabskip\z@{##}$&\global\@eqcnt\@ne
\hskip2\arraycolsep\hfil${##}$\hfil& \global\@eqcnt\tw@\hskip2\arraycolsep
$\displaystyle\tabskip\z@{##}$\hfil
\tabskip\@centering&\llap{##}\tabskip\z@\cr}
\def\endeqnsystem{\@@eqncr\egroup$$\global\@ignoretrue} \makeatother

\begin{document}
\begin{flushright}
\footnotesize
%{CERN-PH-TH/2009-xxx}
%{SACLAY--T15/xxx}
\end{flushright}
\color{black}
%\vspace{0.3cm}

\begin{center}
{\Huge\bf {\sc Integral} X-ray constraints\\[3mm] on sub-GeV Dark Matter}

\medskip
\bigskip\color{black}\vspace{0.6cm}

{
{\large\bf Marco Cirelli}\ $^a$,
{\large\bf Nicolao Fornengo}\ $^b$,\\[3mm]
{\large\bf Bradley J.~Kavanagh}\ $^{c}$,
{\large\bf Elena Pinetti}\ $^{a,b}$
}
\\[7mm]
%{\it $^a$ CERN Theory Division, CERN, \\ 
%Case C01600, CH-1211 Gen\`eve, Switzerland}\\[3mm]	
{\it $^a$ \href{http://www.lpthe.jussieu.fr/spip/index.php}{Laboratoire de Physique Th\'eorique et Hautes Energies (LPTHE)},\\ UMR 7589 CNRS \& Sorbonne University, 4 Place Jussieu, F-75252, Paris, France}\\[3mm]
%{\it $^b$ \href{http://www.iap.fr}{Institut d'Astrophysique de Paris}, \\ UMR 7095 CNRS, Universit\'e Pierre et Marie Curie, \\
%98 bis Boulevard Arago, Paris 75014, France}
{\it $^b$ \href{http://www.df.unito.it/do/home.pl}{Dipartimento di Fisica}, Universit\`a di Torino \& INFN, Sezione\ di Torino,\\ via P. Giuria 1, I-10125 Torino, Italy}\\[3mm]
{\it $^c$ \href{https://ifca.unican.es/en-us}{Instituto de F\'isica de Cantabria}, (IFCA, UC-CSIC),\\ Av. de Los Castros s/n, 39005 Santander, Spain}\\%[3mm]
%{\it $^d$ \href{https://iop.fnwi.uva.nl/grappa/}{Gravitation Astroparticle Physics Amsterdam (GRAPPA)}, \\ Institute for Theoretical Physics Amsterdam \& Delta Institute for Theoretical Physics, \\
%University of Amsterdam, Science Park 904, 1098 XH Amsterdam, The Netherlands}\\
\end{center}

\bigskip

\centerline{\large\bf Abstract}
\begin{quote}
\large
Light Dark Matter (DM), defined here as having a mass between 1 MeV and about 1 GeV, is an interesting possibility both theoretically and phenomenologically, at one of the frontiers of current progress in the field of DM searches. Its indirect detection via gamma-rays is challenged by the scarcity of experiments in the MeV-GeV region. We look therefore at lower-energy X-ray data from the {\sc Integral} telescope, and compare them with the predicted DM flux. We derive bounds which are competitive with existing ones from other techniques. Crucially, we include the contribution from inverse Compton scattering on galactic radiation fields and the CMB, which leads to much stronger constraints than in previous studies for DM masses above 20 MeV.
\end{quote}

\tableofcontents
\newpage

\section{Introduction}
\label{sec:introduction}

The possibility that Dark Matter (DM) consists of a {\em light} particle has gained increasing attention recently (for definiteness we intend here a mass between a few MeV and about a GeV, as we discuss below). Searches for DM have long been dominated by the paradigm of (heavier) WIMPs \cite{Jungman:1995df,Arcadi:2017kky}, but with no convincing WIMP signal observed so far in Direct Detection \cite{Schumann:2019eaa}%Liu:2017drf}
, Indirect Detection \cite{Cirelli:2015gux,Gaskins:2016cha,Hooper:2018kfv} 
or Collider searches \cite{Buchmueller:2017qhf,Kahlhoefer:2017dnp}, 
the attention is turning to lighter (or heavier) candidates. 
Light DM is indeed in some sense a new frontier for searches, requiring new analysis strategies and experimental techniques to achieve sensitivity~\cite{Knapen:2017xzo}.

\medskip

In Direct Detection (DD), most current experiments lose sensitivity for DM masses below  $\sim1$~GeV. This is because the standard method of detection relies on detecting the small amounts of energy deposited by DM via nuclear recoils, which becomes ineffective for DM much lighter than a typical nucleus. However, many significant efforts are underway to explore the sub-GeV regime. This includes extending the sensitivity of `traditional' nuclear recoil detectors to ultra-low energy thresholds \cite{Kouvaris:2016afs,Davis:2017noy} 
({\sc Xenon-10}~\cite{Essig:2012yx}, 
{\sc Xenon-1t}~\cite{Aprile:2019xxb},
{\sc Lux}~\cite{McCabe:2017rln,Akerib:2018hck},
{\sc Cresst}~\cite{Angloher:2015ewa,Angloher:2017sxg,Petricca:2017zdp,Abdelhameed:2019szb}, 
{\sc Super-CDMS}~\cite{Agnese:2016cpb},
{\sc News-G}~\cite{Arnaud:2017bjh},
{\sc DarkSide}~\cite{Agnes:2018ves},
{\sc Edelweiss}~\cite{Armengaud:2019kfj} and in the future 
{\sc Lbeca}~\cite{Bernstein:2020cpc}), 
possibly exploiting the production of detectable signals via DM-electron scattering~\cite{Essig:2011nj,Lee:2015qva,Essig:2017kqs,Derenzo:2016fse}, 
the Migdal effect~\footnote{In the standard DD method, the ionization signal is typically produced as the nucleus struck by the DM particle recoils and hits in turn other nuclei, freeing their electrons. This requires a certain non-negligible energy deposition. The Migdal effect \cite{Migdal1941, Ibe:2017yqa, Vergados:2004bm, Moustakidis:2005gx, Dolan:2017xbu, Baxter:2019pnz,GrillidiCortona:2020owp} refers instead to the fact that even a slowly recoiling or just shaken nucleus can lose one or some of its electrons, and therefore produce a signal. Standard ionization is an in-medium process, while the Migdal effect can take place even for the scattering of an isolated atom.} \cite{Bernabei:2007jz,Akerib:2018hck,Essig:2019xkx} 
or the interactions of DM with the Sun or with Cosmic Rays~\cite{An:2017ojc,Emken:2017hnp,Cappiello:2018hsu,Bringmann:2018cvk,Ema:2018bih,Yin:2018yjn,Cappiello:2019qsw,Alvey:2019zaa,Wang:2019jtk,Guo:2020drq,Kannike:2020agf,Fornal:2020npv,Su:2020zny}. 
Efforts are underway as well to develop new detection strategies~\cite{Griffin:2019mvc,Trickle:2019nya}, 
including the use of semiconductors~\cite{Graham:2012su,Essig:2015cda,Andersson:2020uwc} (such as in the {\sc Sensei} \cite{Tiffenberg:2017aac} and {\sc Damic} \cite{Aguilar-Arevalo:2019wdi,Castello-Mor:2020jhd} experiments), 
superconductors~\cite{Hochberg:2015pha,Hochberg:2015fth}, 
superfluid helium~\cite{Guo:2013dt,Knapen:2016cue,Hertel:2018aal,Acanfora:2019con,Caputo:2019cyg,Caputo:2019xum}, 
evaporating helium~\cite{Maris:2017xvi}, 
2D materials such as arrays of nanotubes and graphene~\cite{Cavoto:2016lqo,Hochberg:2016ntt}, 
1D materials such as superconducting nanowires~\cite{Hochberg:2019cyy},
chemical bond breaking~\cite{Essig:2016crl}, 
production of color-center defects in crystals~\cite{Budnik:2017sbu}, 
polar materials (i.e.~crystals with easily excitable polar atomic bonds~\cite{Knapen:2017ekk,Griffin:2018bjn}),
paleo-detectors~\cite{Baum:2018tfw,Drukier:2018pdy}, 
magnetic bubble chambers~\cite{Bunting:2017net}, 
Casimir forces between nucleons~\cite{Fichet:2017bng},
aromatic targets~\cite{Blanco:2019lrf},
molecular gas excitations~\cite{Essig:2019kfe} or
diamonds~\cite{Kurinsky:2019pgb}. 
Such a long list illustrates the strong interest in the DD community for this regime, and bears well for the future.
  
On the side of Collider searches, the situation has some resemblance with the DD case. The current experiments and flagship colliders (essentially the {\sc Lhc}) are not suitable for exploring the low mass region: the staple signature of DM production is the missing energy corresponding to the DM mass, which in the sub-GeV case is swamped in the experimental background. The strategy is therefore to turn to the search for associated states, i.e.~particles that belong to a new `dark sector', more extended than just the DM particle. In particular, if sub-GeV DM is to be produced by thermal freeze-out in the Early Universe, this requires the existence of new light force mediators (`dark photons'). These mediators provide a link between the dark and visible sectors and are therefore subject to active searches.
Indeed many theory analyses \cite{Borodatchenkova:2005ct,Batell:2009di,deNiverville:2011it,deNiverville:2012ij,Essig:2013vha,Dobrescu:2014ita} and experimental projects \cite{Alekhin:2015byh,Berlin:2018bsc,Akesson:2018vlm,Doria:2018sfx} pursue this direction.
For a detailed and recent review, see e.g. Ref.~\cite{Battaglieri:2017aum}, where additional references can be found.   

In Indirect Detection (ID), one typically searches for the Standard Model particles (charged particles, such as electrons and positrons, or  neutral ones, such as gamma rays and neutrinos) produced in the annihilation of DM in the Galaxy, with energies at or just below the DM mass. Concerning charged particles, the problem is that solar activity holds back sub-GeV charged cosmic rays and therefore we have no access to them.\footnote{An exception to this point is the use of data from the {\sc Voyager} spacecraft, which is making measurements outside of the heliosphere~\cite{Boudaud:2016mos}. We will comment on the corresponding constraints later.} Concerning gamma-rays, the sensitivity of the most powerful of the recent telescopes, {\sc Fermi-Lat}, stops at about 100 MeV, so that typically DM masses only as low as about 1 GeV can be probed. At much lower energies, below a few MeV, one has competitive data from {\sc Integral}. But in-between $\sim$1 and 250 MeV, only relatively old data from {\sc Comptel} are available and no current competitive experiment exists. Indeed, a number of authors have discussed proposals to fill this so-called `MeV gap' in a useful way for DM searches~\cite{Beacom:2004pe,Rasera:2005sa,Boggs:2006gi,Greiner:2011ih,Bandstra:2011pk,Hunter:2013wla,Wu:2014tya,Boddy:2015efa,Bartels:2017dpb,Gonzalez-Morales:2017jkx,DeAngelis:2016slk,Kierans:2017bmv,Kumar:2018heq}. 
Alternatively, low energy neutrinos have been considered, but they are found not to be very competitive with respect to photons (e.g.~the projected sensitivity of 20 years of run with the future {\sc HyperKamiokaNDE} detector is weaker than the existing bound that we will discuss below, and the new ones we will derive)~\cite{PalomaresRuiz:2007eu,Bell:2020rkw}.

\medskip

Another possibility for ID of such light DM (which is the one we will entertain here) is to look at a range of energies much lower than that of the DM mass.\footnote{For former applications of the same idea to heavy, WIMP-like, DM see for instance \cite{Colafrancesco:2005ji,Colafrancesco:2006he,Cholis:2008wq,Zhang:2008tb,Borriello:2009fa,Barger:2009yt,Cirelli:2009vg,Cirelli:2009dv,Zavala:2011tt,Beck:2015rna}.} The basic idea is the following: the electrons and positrons produced in the Galactic halo by the annihilations of DM particles with a mass $m\simeq 1$ GeV have naturally an energy $E \lesssim 1$ GeV; they undergo Inverse Compton scattering (ICS) on the low energy photons of the ambient bath (the CMB, infrared light and starlight) and produce X-rays, which can be searched for in X-ray surveys. Indeed, the ICS process increases the photon energy from the initial low value $E_0$ to a final value $E \approx 4 \gamma^2 E_0$ upon scattering off an electron with relativistic factor $\gamma = E_e/m_e$. Hence, a 1 GeV electron will produce a $\sim$ 1.5 keV X-ray when scattering off the CMB ($E_0 \approx 10^{-4}$ eV). By the same token, a mildly-relativistic MeV electron will produce a $\sim$ 0.15 keV X-ray when scattering off UV starlight ($E_0 \approx 10$ eV). These considerations roughly define our range of interest for DM masses: $m_{\rm DM} \simeq$ 1 MeV $\to$ 1 GeV.\footnote{Note that we are not interested in keV DM, that can e.g. produce X-rays by direct annihilation or decay. That is a whole other set of searches, e.g.~for keV sterile neutrino DM~\cite{Bulbul:2014sua,Boyarsky:2014jta}.} 
So our goal is to explore whether there are cases in which X-ray observations can impose constraints on  sub-GeV DM that would otherwise fall below the sensitivity of the more conventional gamma-ray searches. To this end, we focus on data from the \textsc{Integral} X-ray satellite~\cite{Bouchet:2011fn}.

\medskip

The electrons and positrons produced by the annihilation of light DM propagate in the Galaxy. The dominant energy loss processes for these particles at low energy are bremsstrahlung, ionization and Coulomb scattering. These therefore reduce the power that is effectively injected into ICS. However these processes are relevant where gas is present, i.e. essentially in the Galactic center and within the disk. We will therefore focus on observational windows at mid-to-high Galactic latitudes, where indeed ICS (together with synchrotron emission) is a relevant mechanism of photon production.

\bigskip

From the theoretical point of view, the case for sub-GeV DM is arguably not as strong as the traditional case for WIMPs. On the other hand, considering the lack of evidence for a DM in the typical WIMP mass range, it is definitely worth leaving no stone unturned~\cite{Bertone:2018krk}. 
From the phenomenological perspective, light (scalar) DM has been invoked as a viable possibility~\cite{Boehm:2002yz,Boehm:2003hm,Fayet:2007ua} and as an explanation~\cite{Boehm:2003bt,Ahn:2005ck,Boehm:2006mi,Ema:2020fit} of the 511 KeV line from the center of the galaxy~\cite{Prantzos:2010wi}.
From the point of view of theoretical motivations, well founded models include for instance the SIMP scenarios~\cite{Hochberg:2014dra,Boddy:2014yra,Hochberg:2014kqa,Choi:2017zww,Berlin:2018tvf}, the WIMPless idea~\cite{Feng:2008ya}, Forbidden Dark Matter~\cite{DAgnolo:2015ujb}, axinos~\cite{Covi:1999ty,Choi:2011yf}, neutrino mass generation~\cite{Boehm:2006mi,Arhrib:2015dez} and even certain SuSy configurations~\cite{Boehm:2003ha,Hooper:2008im,Essig:2010ye}: all these constructions naturally include or even predict sub-GeV DM particles.
One may also add to the list asymmetric DM~\cite{Petraki:2013wwa}: in the standard lore it naturally predicts DM with a mass of a few GeV and no annihilation signals, but actually the mass can be lighter~\cite{Falkowski:2011xh,Lin:2011gj} and residual or late annihilations can occur and give rise to an interesting phenomenology~\cite{Falkowski:2011xh,Baldes:2017gzu}.  Freeze-in DM production~\cite{Hall:2009bx} is also often cited in connection to sub-GeV DM, not necessarily because freeze-in points to the sub-GeV range but because it provides a viable DM production mechanism.

More broadly, studies addressing the motivations, the phenomenology and the constraints (notably from cosmology) of MeV-GeV DM include \cite{Serpico:2004nm,Hooper:2007tu,Krnjaic:2015mbs,Wilkinson:2016gsy,Bertuzzo:2017lwt,Hufnagel:2017dgo,Darme:2017glc,Xu:2018efh,Dutra:2018gmv,Hufnagel:2018bjp,Forestell:2018txr,Choudhury:2019tss,Choudhury:2019sxt,Sabti:2019mhn,Bondarenko:2019vrb,Coffey:2020oir,Katz:2020ywn}.

\medskip

The rest of this paper is organized as follows. 
In Sec.~\ref{sec:computations} we review the computation of X-rays from DM annihilations, via the ICS process. 
In Sec.~\ref{sec:data} we present the X-rays measurements by {\sc Integral} that we employ. 
In Sec.~\ref{sec:results} we present the results, in particular deriving constraints in the usual plane of annihilation cross section versus mass of the DM particle, and we briefly compare with other existing constraints.
In Sec.~\ref{sec:conclusions} we summarize and conclude.

%%%%%%%%%%%%%%%%%%%%%%%%%%%%%%%%%%%%%%%%%%%%%%%%%%%%%%%%

\section{X-rays from DM annihilations}
\label{sec:computations}

In this section we briefly review  the basic formalism for (soft $\gamma$-ray and) hard X-ray production from DM annihilations, essentially via final state radiation and ICS emission. For all details we refer the reader to Ref.~\cite{Cirelli:2010xx,Buch:2015iya}, whose notation we mostly follow. Two important differences however apply. First, Ref.~\cite{Cirelli:2010xx,Buch:2015iya} do not deal with DM masses smaller than 5 GeV, so that we cannot straightforwardly use the tools developed there. Second, in Ref.~\cite{Cirelli:2010xx,Buch:2015iya} a detailed treatment of the energy-loss and diffusion process is employed. Since, as mentioned above, we are mostly interested in higher latitude signals where ICS energy losses dominate, we adopt a simplified treatment (described below).

\medskip

Let us consider a direction of observation from Earth that is identified by the angle $\theta$ (the aperture between the direction of observation and the axis connecting the Earth to the Galactic Center), or, equivalently, by the latitude and longitude pair $(b,\ell)$. At each point along this direction, DM particles annihilate, contributing to the photon signal, which we collect by integrating all the contributions along the line of sight. Since we focus on DM lighter than a few GeV, we consider only three annihilation channels:
\begin{align}
{\rm DM \, DM} &\to e^+e^-,  \label{eq:ee}\\
{\rm DM \, DM} &\to \mu^+\mu^-, \label{eq:mumu}\\
{\rm DM \, DM} &\to \pi^+\pi^-, \label{eq:pipi}
\end{align}
which are kinematically open whenever $m_{\rm DM} > m_i$ (with $i=e,\mu,\pi$) and that we study  one at a time. The pion channel is representative of a hadronic DM channel. We do not consider the annihilation into a pair of neutral pions, since in this case the (boosted to the DM frame) $\gamma$-rays do not reach down to the energies covered by {\sc Integral}.

For each channel, the total photon flux is given by the sum of two contributions: the emission from the charged particles in the final state (from Final State Radiation, FSR, and, whenever relevant, other radiative decays, Rad) and the photons produced via ICS by DM-produced energetic $e^\pm$. Also, another contribution to the total photon flux is given by the in-flight annihilation of DM-produced $e^+$ with ambient $e^-$ from the gas~\cite{Beacom:2005qv}. This process is however subdominant outside of the gas-dense region of the GC  that we do not consider, as mentioned above.

\medskip

The differential flux of the {\em Final State Radiation} or {\em Radiative Decays} photons from the annihilations of a DM particle of mass $m_{\rm DM}$ is computed via the standard expression (see e.g.~Ref.~\cite{Slatyer:2017sev}):
\beq
\label{eq:FSRflux}
\frac{d \Phi_{{\rm FSR}\gamma}}{dE_\gamma \, d\Omega} = \frac{1}{2}\frac{r_\odot}{4\pi} \left(\frac{\rho_\odot}{m_{\rm DM}}\right)^2 J(\theta) \
\langle \sigma v\rangle_f \frac{dN_{{\rm FSR} \gamma}^f}{dE_\gamma}  ,\qquad
J(\theta) = \int_{\rm l.o.s.} \frac{ds}{r_\odot} \left(\frac{\rho(r(s,\theta))}{\rho_\odot}\right)^2.
\eeq
where $r_\odot \simeq 8.33$ kpc and $\rho_\odot = 0.3$ GeV/cm$^3$ are the conventional values for the distance of the Sun from the Galactic center (GC) and the local DM density.
Here, $E_\gamma$ denotes the photon energy,  $d\Omega$ is the solid angle, $r$ is the distance from the GC  in spherical coordinates, in turn expressed in terms of $s$, the coordinate that runs along the line of sight ($s=0$ corresponds to the Earth position), and the angle $\theta$: $r^2 = s^2+r_\odot^2-2 s r_\odot\cos\theta$. We use natural units $c = \hbar = 1$ throughout the paper. The normalized $J$ factor corresponds to the integration along the line of sight of the square of the galactic DM density profile $\rho(r)$, for which we adopt a standard Navarro-Frenk-White profile \cite{Navarro:1995iw,Cirelli:2010xx}
\beq
\rho_{\rm NFW}(r) = \rho_{s}\frac{r_{s}}{r}\left(1+\frac{r}{r_{s}}\right)^{-2} \qquad {\rm with } \ \rho_{\rm s} = 0.184 \ {\rm GeV/cm}^3, \quad r_{\rm s} = 24.42 \ {\rm kpc}.
\eeq
The annihilation cross section in each of the final states $f$ of Eqs.~(\ref{eq:ee})$-$(\ref{eq:pipi}) is denoted $\langle \sigma v\rangle_f$. 

\smallskip

For the spectrum of FSR photons we adopt the following expressions which we derive from the analysis of\cite{Bystritskiy:2005ib} and reproduce here for the convenience of the reader:
\beq
\label{eq:FSRll}
\frac{dN_{{\rm FSR} \gamma}^{l^+l^-}}{dE_\gamma} = \frac{\alpha}{\pi\beta(3-\beta^2)\,m_{\rm DM}} 
\left[{\cal{A}} \, \ln \frac{1+R(\nu)}{1-R(\nu)}-2\, {\cal{B}}\,  R(\nu)
\right] \,,
\eeq
where
\beq
{\cal{A}} = \left[\frac{\left(1+\beta^{2}\right)\left(3-\beta^{2}\right)}{\nu}-2\left(3-\beta^{2}\right)+2 \nu\right] \,,
\eeq
\beq
{\cal{B}} = \left[\frac{3-\beta^{2}}{\nu}(1-\nu)+\nu\right] \,,
\eeq
where $l=e,\mu$ and we have defined: $\nu = E_\gamma/m_{\rm DM}$,  $\beta^2 = 1-4\mu^2$ with $\mu=m_l/(2\,m_{\rm DM})$ and $R(\nu) = \sqrt{1-4\mu^2/(1-\nu)}$.  For the pion \cite{Bystritskiy:2005ib}
\beq
\label{eq:FSRpipi}
\frac{dN_{{\rm FSR} \gamma}^{\pi^+\pi^-}}{dE_\gamma} = 
\frac{2 \alpha}{\pi\beta\,m_{\rm DM}}\left[\left(\frac{\nu}{\beta^{2}}-\frac{1-\nu}{\nu}\right) R(\nu)+\left(\frac{1+\beta^{2}}{2 \nu}-1\right) \ln \frac{1+R(\nu)}{1-R(\nu)}\right] \,,
\eeq
with the same definitions as above with $m_l \rightarrow m_\pi$.

\smallskip

The muon can undergo the radiative decay $\mu^- \rightarrow e^- \, \bar{\nu}_e \,  \nu_\mu \, \gamma$, and this channel can be a relevant source of soft photons. For the photon spectrum we adopt the parameterisation of  Ref. \cite{Essig:2009jx, Coogan:2019qpu} which in turn are derived from Ref. \cite{Kuno:1999jp}. In the muon rest frame the expression we adopt is therefore
\beq
\left.\frac{dN_{{\rm Rad}\, \gamma}^{\mu}}{dE_\gamma} \right|_{E_\mu = m_\mu} = \frac{\alpha(1-x)}{36 \pi E_{\gamma}}\left[12\left(3-2 x(1-x)^{2}\right) \log \left(\frac{1-x}{r}\right) +x(1-x)(46-55 x)-102 \right]\,,
\label{eq:muon_rad}
\eeq
where $x = 2E_\gamma/m_\mu$, $r = (m_e/m_\mu)^2$ and the maximal photon energy is $E^{\rm max}_\gamma = m_\mu (1-r)/2 \simeq 52.8$ MeV. For muons in flight,  Eq.~(\ref{eq:muon_rad}) is boosted to the frame where the muon has energy $E_\mu = m_{\rm DM}$.  For the process ${\rm DM} \, {\rm DM} \rightarrow \mu^+ \mu^-$, a multiplicity factor of 2 needs to be applied, since a pair of muons is produced for each annihilation event.  

Charged pions can also produce photons radiatively through the process $\pi^- \rightarrow \ell^- \, \bar{\nu}_\ell\, \gamma$ with $\ell=e,\mu$. Also in this case we adopt the parameterisation of  Ref. \cite{Coogan:2019qpu} which has been derived from Ref. \cite{Bryman:1982et}. In the pion rest frame the expression we adopt is then
\beq
\left.\frac{dN_{{\rm Rad}\, \gamma}^{\pi}}{dE_\gamma}\right|_{E_\pi = m_\pi} = \frac{\alpha[f(x)+g(x)]}{24\, \pi \,m_{\pi} \,f_{\pi}^{2}\,(r-1)^{2}\,(x-1)^{2} \,r \,x}\,,
\eeq
where $x = 2E_\gamma/m_\pi$, $r = (m_\ell/m_\pi)^2$, $f_\pi = 92.2$ MeV is the pion decay constant and
\begin{eqnarray}
f(x)  &=& (r+x-1)\left[m_{\pi}^{2}\, x^{4}\left(F_{A}^{2}+F_{V}^{2}\right)\left(r^{2}-r x+r-2(x-1)^{2}\right)\right. \nonumber\\
&& -12 \,\sqrt{2} \,f_{\pi}\, m_{\pi} \,r(x-1) x^{2}\left(F_{A}(r-2 x+1)+x F_{V}\right) \nonumber\\
&& \left.-24 \,f_{\pi}^{2}\, r(x-1)\left(4 r(x-1)+(x-2)^{2}\right)\right] \,,\\
g(x) &=& 12\, \sqrt{2} \, f_\pi \, r(x-1)^2 \log\left(\frac{r}{1-x}\right) [m_{\pi}\, x^{2} (F_{A}\, (x-2r) - x \, F_{V}) \nonumber \\
&&+\sqrt{2}\, f_{\pi} (2 r^{2}-2 r x-x^{2}+2 x-2)] \,,\nonumber
\end{eqnarray}
where the axial and vector form factors are $F_{A}=0.0119$ and $F_{V}\left(q^{2}\right)=F_{V}(0)\left(1+a q^{2}\right)$ with $F_{V}(0)=0.0254$, $a=0.10$, $q^2 = (1-x)$ \cite{Coogan:2019qpu, Tanabashi:2018oca}.

When the pion decays into an on-shell muon, the radiative decay of the muon is again a source of low energy photons. Following Refs. \cite{Coogan:2019qpu}, the total radiative charged pion spectrum in the pion rest frame can be phrased as
\beq
\left.\frac{dN_{{\rm RadTot}\, \gamma}^{\pi}}{dE_\gamma}\right|_{E_\pi = m_\pi} =
\sum_{\ell=e, \mu} {\rm BR}\left(\pi \rightarrow \ell \nu_{\ell}\right) \,
\left.\frac{dN_{{\rm Rad}\, \gamma}^{\pi}}{dE_\gamma}\right|_{E_\pi = m_\pi}
+{\rm BR}(\pi \rightarrow \mu \nu_{\mu}) 
\left.\frac{dN_{{\rm Rad}\, \gamma}^{\mu}}{dE_\gamma}\right|_{E_\mu = E_\star}
\label{eq:pion_radtot}
\eeq
where $E_\star=(m_\pi^{2}+m_{\mu}^{2})/(2 m_\pi)$ is the muon energy in the pion rest frame.
For the process ${\rm DM} \, {\rm DM} \rightarrow \pi^+ \pi^-$, where the pion is in flight,  Eq.~(\ref{eq:pion_radtot}) is  boosted to the frame where $E_\pi = m_{\rm DM}$ and a multiplicity factor of 2 is applied, since also in this case a pair of pions is produced for each annihilation event.
As pointed out in Ref. \cite{Coogan:2019qpu}, the muon radiative decay provides a quite relevant soft-photon production channel. Both in the muon  and in the pion annihilation channel, the Rad photon flux arising from the muon radiative decay is dominant over the FSR emission for DM masses up to a few GeV.  This will be explicitly seen in the derivation on the DM annihilation cross section bounds in Section \ref{sec:results}.

\bigskip

\begin{figure}[t]
\begin{center}
\includegraphics[width= 0.48 \textwidth]{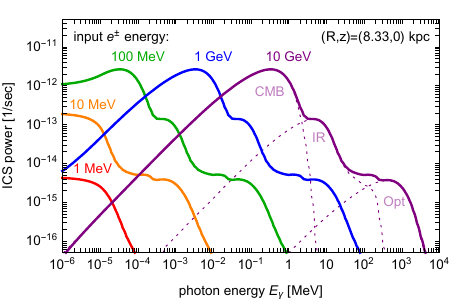} \quad
\includegraphics[width= 0.2615 \textwidth]{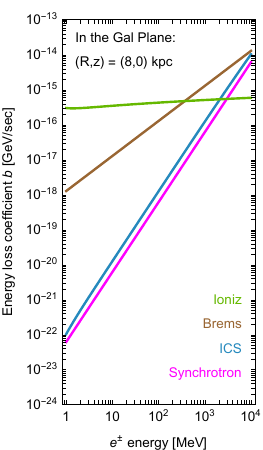} \hspace{-0.6cm}
\includegraphics[width= 0.23 \textwidth]{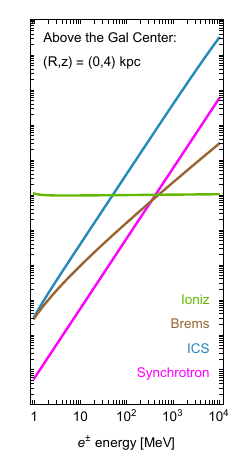}
\caption{\em \small \label{fig:powerandB} {\em Left:} {\bfseries Inverse Compton Scattering (ICS) power} as a function of the emitted photon energy, for the listed input $e^\pm$ energies. In the 10 GeV case, for illustration we show the contributions of the three components of the photon bath: CMB, infrared light and optical light. {\em Right:} {\bfseries Energy losses} for electrons and positrons in the energy regime of interest. The first subpanel refers to the situation in the Galactic Plane, in a spot similar to the location of the solar system: one sees that the $e^\pm$ lose most of their energy via interactions with the gas (ionization and bremsstrahlung). The second subpanel is for a point well outside the plane, on the vertical of the Galactic Center: one sees that the ICS emission is dominant down to $\sim$40 MeV. Lines of sight that avoid the Galactic Plane (i.e.~at high latitude) are therefore preferred for our purposes.}
\end{center}
\end{figure}

The differential flux of the {\em Inverse Compton Scattering photons} received at Earth is written in terms of the emissivities $j(E_\gamma,\vec x)$ of all cells located along the line of sight at position $\vec x$: 
\beq
\label{eq:ICSflux}
\frac{d\Phi_{{\rm IC}\gamma}}{dE_\gamma \, d\Omega} = \frac 1{E_\gamma} \int_{\rm l.o.s.} ds\, \frac{j(E_\gamma,\vec x(s,b,\ell))}{4\pi}.
\eeq
Here the spherical symmetry of the system around the center of the Galaxy is broken by the distribution of the ambient light, which mostly lies in the galactic disk. Hence we use the 3D notation $\vec x$ instead of $r$, and we need the $(b,\ell)$ coordinates instead of the single angle $\theta$. These quantities are related by the usual expression $\cos\theta = \cos b\cos l$. It will also be convenient to work in cylindrical coordinates $(R,z)$.\\
The emissivity is obtained as a convolution of the density of the emitting medium with the power that it radiates. In this case therefore 
\beq
\label{Rademissivity}
j(E_\gamma,\vec x)=2\int_{m_e}^{M_{\rm DM}}dE_e\ \mathcal{P}_{\rm IC}(E_\gamma,E_e,\vec x)\ \frac{dn_{e^\pm}}{dE_e}(E_e,\vec x),
\eeq
where $E_e$ denotes the electron total energy and $dn_{e^\pm}/dE_e$ is the local electron (or positron) spectral number density (the overall factor of 2 takes into account the equal populations of electrons and positrons). 
$\mathcal P=\sum_i \mathcal P_{\rm IC}^i$ is the differential power emitted into photons due to ICS radiative processes 
\beq
\mathcal P_{\rm IC}^i(E_\gamma, E_e,\vec  x) = E_\gamma \int d\epsilon \ n_i(\epsilon,\vec x) \ \sigma_{\rm IC}(\epsilon, E_\gamma, E_e)\,,
\eeq
where the sum on $i$ runs over the different components of the photon bath, expressed by their photon number densities $n_i$ ($\epsilon$ is the photon energy before the scattering): CMB, dust-rescattered infrared light (IR) and optical starlight (SL). For the last two components, we use the Interstellar Radiation Field (ISRF) maps extracted from \cite{Vladimirov:2010aq}, as in \cite{Cirelli:2009vg,Cirelli:2010xx,Buch:2015iya}. We do not detail here further the computations (we refer the reader to~\cite{Petruk:2008bb,Cirelli:2009vg,Cirelli:2010xx,Buch:2015iya} and references therein) but, for reference, we plot the total power $\mathcal P$ in the energy regime of interest in Fig.~\ref{fig:powerandB}, left. In the above equation, $\sigma_{\rm IC}$ is the Klein-Nishina cross section.

\bigskip

The local $e^\pm$ spectral number density we are concerned with here is the result of injection due to DM annihilation, plus the subsequent diffusion of the $e^\pm$ in the local Galactic environment, subject to energy losses by several processes (see below). 
 We adopt a simplified treatment by neglecting diffusion, i.e.~we assume that electrons and positrons scatter off the ambient photons in the same location where these $e^\pm$ were produced by annihilation events of DM particles.
This is sometimes referred to as the `on the spot' approximation. While this is certainly a drastic approximation, we verified a posteriori that it is justified in the regimes of our interest: the timescales of diffusion over lengths comparable to the angular bins of the {\sc Integral} data are longer than the loss timescales, except for lines of sight towards high latitudes, which however contribute little to the signal.
In this regime, the quantity can be simply expressed as (see e.g.~\cite{Cirelli:2009vg})
\beq
\frac{dn_{e^\pm}}{dE_e}(E_e,\vec x) = \frac1{b_{\rm tot}(E_e,\vec x)}\int_{E_e}^{m_{\rm DM}}d\tilde E_e\ Q_e(\tilde E_e, \vec x), \quad {\rm with} \quad Q_e(\tilde E_e,\vec  x) = \frac{\langle \sigma v\rangle}{2} \left( \frac{\rho(\vec x)}{m_{\rm DM}} \right)^2 \frac{dN_{e^\pm}}{d \tilde E_e}.
\eeq
Here $b_{\rm tot}(E,\vec x) \equiv -\frac{{\rm d}E}{{\rm d}t} = b_{\rm Coul+ioniz} + b_{\rm brem} +  b_{\rm syn} + b_{\rm ICS}$ is the energy loss function, which takes  into account all the energy loss processes that the $e^\pm$ suffer in the local Galactic environment in which they are injected. These are due (in order of importance for increasing $e^\pm$ energy: Coulomb interactions and ionization, bremsstrahlung, synchrotron emission and Inverse Compton scattering). We plot the energy loss function in Fig.~\ref{fig:powerandB} (right) for illustration and we refer the reader to \cite{Buch:2015iya} for more details. 
 It is important to stress that these quantities are subject to astrophysical uncertainties, stemming e.g.~from the uncertainties in the gas distribution, in the ISRF and in the intensity of the galactic magnetic field. We will discuss the impact of the such uncertainties on our signal at the end of this Section.
$Q_e(\tilde E_e, \vec x)$ is the injection term of electrons and positrons from DM annihilations: its integral over $\tilde E_e$ corresponds essentially to the number of electrons generated with an energy larger than $E_e$ at position $\vec x$. We cut the integral in the horizontal direction at $R = R_{\rm gal} = 20$ kpc, where $R_{\rm gal}$ is the presumed radial extension of the Milky Way, and in the vertical direction at $z = 4$ kpc, assuming this to be the size of the magnetic halo that keeps the $e^\pm$ confined.

\bigskip

The last ingredient needed, the $e^\pm$ spectrum from DM annihilations in the different cases, is rather straightforwardly computed. For the $e^+e^-$ channel it consists simply in a monochromatic line of $e^\pm$ with $E_e = m_{\rm DM}$. For the $\mu^+\mu^-$ case, we boost to the DM annihilation frame (where the muon has energy $E_\mu = m_{\rm DM}$) the electron spectrum from muon decay obtained in the muon rest frame \cite{Michel:1950}:
\beq
\frac{d N^{\mu\rightarrow e\nu\bar{\nu}}_e}{d E_{e}}=\frac{4 \sqrt{\xi^{2}-4 \varrho^{2}}}{m_{\mu}}\left[\xi(3-2 \xi) + \varrho^{2}(3 \xi-4)\right]
\label{eq:mudecay}
\eeq
where $\varrho=m_e/m_\mu$, $\xi = 2E_e/m_\mu$ and the maximal electron energy is $E^{\rm max}_e = (m_\mu^2+m_e^2)/(2m_\mu)$. For the $\pi^+\pi^-$ case, due to the decay chain $\pi \rightarrow \mu \rightarrow e$, we  boost the electron spectrum from muon decay of Eq.~(\ref{eq:mudecay}) to the rest frame of the pion (where the muon has energy $E_\mu = (m_\pi^2+m_\mu^2)/(2m_\pi)$) and then boost the ensuing distribution to the DM annihilation frame (where the pion has energy $E_\pi = m_{\rm DM}$). The expression of the boost from a frame where the particle $a$ (produced by its parent $A$) has energy $E'$ and momentum $p'$ to a frame where the parent particle has energy $E_A$
 \cite{Donato:2003xg} reads 
\beq
\frac{dN}{dE} = \frac{1}{2\beta\gamma} \; \int_{E'_{\rm min}}^{E'_{\rm max}} \frac{1}{p'}\;\frac{dN}{dE'}
\eeq
where $\gamma = E_A/m_A$ and $\beta = (1-\gamma^{-2})^{1/2}$ are the Lorentz factors for the boost, and $E'_{\rm max|min} = \gamma (E\pm \beta\,p)$ and the signs $+$ ($-$) refers to max (min), respectively.  When $a$ is produced as monochromatic, i.e. $dN/dE' = \delta(E - E_\star)$, the above expression reduces to the typical box spectrum $dN/dE = 1/(2\beta\gamma p_\star)$ for $\gamma( E_\star - \beta p_\star) \leq E \leq \gamma( E_\star + \beta p_\star) $.

We remind the reader here that our $\pi^+\pi^-$ case is just intended as one possible representative case of DM annihilations into light quarks. A more detailed analysis would require using the spectra computed in \cite{Plehn:2019jeo}, that consider the annihilation into light quark pairs and the subsequent production of many light hadronic resonances besides pions. That would require choosing a specific model for the annihilation diagram (e.g.\ via a light mediator with arbitrary but defined couplings to quarks). In order to keep our analysis as simple and as model-independent as possible, we stick to the case of direct annihilation into charged pions only, deferring a more complete analysis to future work.

\bigskip

\begin{figure}[t]
\begin{center}
\includegraphics[width= 0.49 \textwidth]{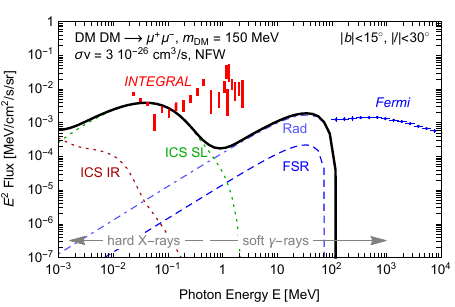}
\includegraphics[width= 0.49 \textwidth]{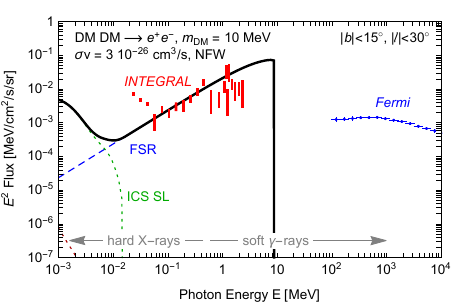}
\caption{\em \small \label{fig:spectra} {\bfseries Example photon spectra from sub-GeV DM, that illustrate our main points}. {\em Left:} 
The hard X-ray and soft $\gamma$-ray spectrum produced by a 150 MeV DM particle annihilating into $\mu^+\mu^-$. We show the different components in color and the total flux in thick black. The spectrum cuts off before reaching {\sc Fermi}'s data (taken from \cite{Ackermann:2014usa} and reported here just for reference, as they are not the focus of our work), but produces a signal in X-rays that can be constrained by {\sc Integral} (taken from \cite{Bouchet:2011fn}, Fig.~7). 
The Final State Radiation (FSR, blue dashed) and Radiative Decay (Rad, blue dot-dashed) contributions yield signals that pass well below the X-ray data. However, the inclusion of the DM-induced Inverse Compton Scattering (ICS) contribution on the different components of the Galactic ambient light (starlight (SL, green dotted), dust-reprocessed infrared light (IR, brown dotted)) and the CMB (not visible in the plot), leads to a flux which is orders of magnitude larger, thus producing stronger constraints.
{\em Right:} The same for a 10 MeV DM particle annihilating into $e^+e^-$. In this case the limit is instead driven by the FSR contribution because the DM ICS contributions fall to too low energy for {\sc Integral}. In these illustrations, the signals are computed over the $|b|<15^\circ, |\ell|<30^\circ$ region of interest (RoI): in our analysis we actually use smaller RoIs, removing low latitudes (see Sec.~\ref{sec:data}).}
\end{center}
\end{figure}

With all the ingredients above, we are therefore in a position to compute the full spectrum of photons from DM annihilation. The final step consists in integrating the contributions in Eqs.~(\ref{eq:FSRflux}) and (\ref{eq:ICSflux}) over the selected region of observation:
\beq
\label{eq:totaldiffflux}
\frac{d\Phi_{{\rm DM}\gamma}}{dE_\gamma}= \int_{b_{\rm min}}^{b_{\rm max}} \int_{\ell_{\rm min}}^{\ell_{\rm max}} db \, d\ell \, \cos b \ \left( \frac{d\Phi_{{\rm FSR}\gamma}}{dE_\gamma \, d\Omega} + \frac{d\Phi_{{\rm IC}\gamma}}{dE_\gamma\, d\Omega} \right).
\eeq
We can then compare the total flux with the data in the same region, in order to derive constraints on Dark Matter. Two examples are presented in Fig.~\ref{fig:spectra}, illustrating the main points of our analysis.

\bigskip

\begin{figure}[!ht]
\begin{center}
	\parbox[b]{.49\textwidth}{
		\includegraphics[width=\linewidth]{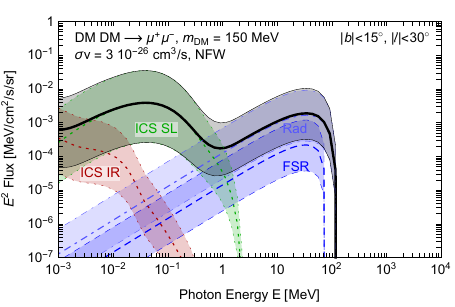}}
	\parbox[b]{.49\textwidth}{
		\caption{\em \small \label{fig:spectra_uncert} {\bfseries Impact of the astrophysical uncertainties} on the same photon spectra of Fig.~\ref{fig:spectra} (left panel). The shaded bands correspond to the cumulative effect of the uncertainty on the DM profile, on the gas density, on the ISRF and on the galactic magnetic field. Each colored band correspond to the different components (blue shade for FSR and Rad, green shade for ICS on StarLight, salmon shade for ICS on InfraRed light).} \vspace{1.cm}
}
\end{center}
\end{figure}

{\bf Astrophysical uncertainties.} As mentioned above, the computed spectra are subject to astrophysical uncertainties of different sorts, that we detail here. First of all, the DM distribution in the Galaxy is uncertain. Choosing a different profile (with respect to the NFW profile specified above) affects all the different components: FSR, Rad and the 3 ICS components. To compute the constraints, we will therefore also adopt a cored profile and a peaked NFW one (characterized by a slope $r^{1.26}$ towards the GC). 
Secondly, the gas density in the Galaxy has significant uncertainties. This affects the energy losses by Coulomb, ionization and bremsstrahlung, therefore in turn affecting the spectrum of the emitting $e^\pm$. We will vary by a factor 2 the overall gas density in the Galaxy.
Thirdly, the ISRF also carries uncertainties, which affect the energy losses by ICS and of course the ICS signal emission. We vary by a factor of 2 overall the intensity of the ISRF in the Galaxy to mimic this error. 
Finally, the galactic magnetic field also carries significant uncertainties, which impact the energy losses by synchrotron. We adopt the different magnetic field configurations discussed in \cite{Buch:2015iya}. The effect however is very limited, since the synchrotron radiation losses are always subdominant in our regime of interest (see Fig.~\ref{fig:powerandB}). Fig.~\ref{fig:spectra_uncert} illustrates the impact of these uncertainties on the spectra.

%%%%%%%%%%%%%%%%%%%%%%%%%%%%%%%%%%%%%%%%%%%%%%%%%%%%%%%%

\section{{\sc Integral} X-ray data and analysis}
\label{sec:data}

\begin{figure}[!t]
	\parbox[b]{.48\linewidth}{
		\includegraphics[width=\linewidth]{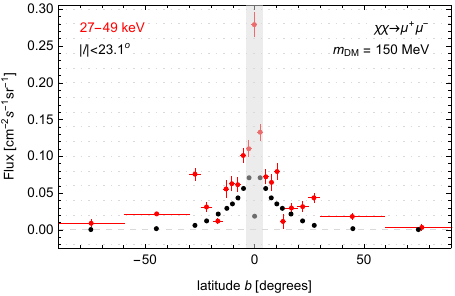}}\hfill
		\vspace{5mm}
	\parbox[b]{.48\linewidth}{ 
		\includegraphics[width=\linewidth]{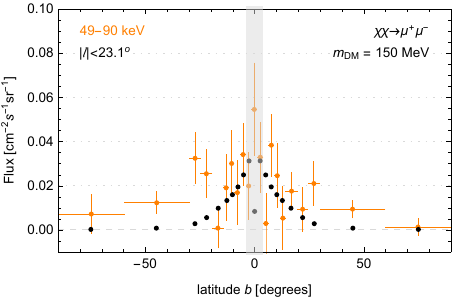}} 
	\parbox[b]{.48\linewidth}{
		\includegraphics[width=\linewidth]{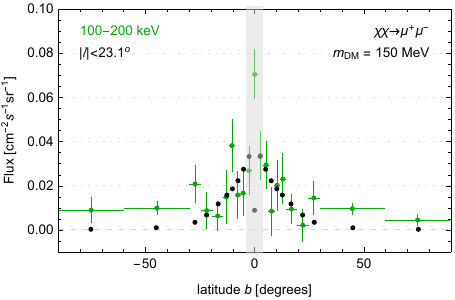}}\hfill
		\vspace{5mm}
	\parbox[b]{.48\linewidth}{ 
		\includegraphics[width=\linewidth]{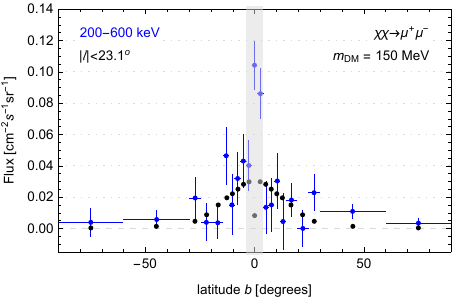}} 
	\parbox[b]{.48\linewidth}{
		\includegraphics[width=\linewidth]{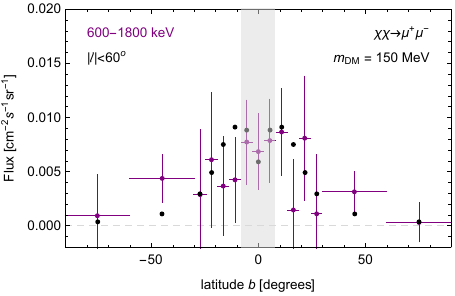}}\hfill
	\parbox[b]{.48\linewidth}{
		\caption{\em \small \label{fig:fluxvsb} {\bfseries {\scshape Integral} data, compared to the maximal predicted DM flux}: in each of the 5 energy bands, we show the data from {\sc Integral} (colored points) as well as the predicted DM flux for 150 MeV DM  annihilating into $\mu^+\mu^-$ (black points). The shades cover the low-latitude points that we do not use. 
		In each panel the annihilation cross section is set at its limit for this DM mass within that specific energy band: i.e.\ it maximizes the flux without exceeding the data in the band by more than allowed by our constraint procedure.} 
}
\end{figure}

We use the data from the {\sc Integral/Spi} X-ray spectrometer, as reported in \cite{Bouchet:2011fn}, which follows previous work in \cite{Bouchet:2008rp,Bouchet:2005ys}. The data were collected in the period 2003$-$2009, corresponding to a significant total exposure of about $10^8$ seconds, and cover a range in energy between 20 KeV and a few MeV. They are provided both in the form of a spectrum of the total diffuse flux in a rectangular region of observation centered around the GC ($|b|<15^\circ, |\ell||<30^\circ$, Fig.~6 and Fig.~7 in \cite{Bouchet:2011fn}) and in the form of an angular flux in latitude and longitude bins (Figs.~4 and 5 in \cite{Bouchet:2011fn}). In Fig.~\ref{fig:spectra}, for illustration, we showed the former set of data. In the following, however, we will use the latter set of data, from which we cut out the Galactic Plane (GP). This is based on two reasons, both of which make the GP less attractive from an analysis perspective.
First, the GP is obviously bright in X-rays due to many astrophysical sources (dust emission, IC emission from cosmic rays, etc.), which have no connection to Dark Matter but represent a significant source of background. In our analysis, in order to  adopt a conservative approach in the derivation of the DM bounds, we will not attempt to model and subtract the galactic X-ray emission and we therefore choose to stay away from the most contaminated region. Second, the $e^\pm$ emissivity in the GP is dominated by scattering processes on gas (Coulomb interactions, ionization and bremsstrahlung), due to its high density there. This complexity of the GP induces uncertainties in the predictions of the photon fluxes, which can be avoided by looking at higher latitudes. Therefore we choose to focus on relatively gas-poorer regions in which the ICS emissivity is the most relevant  process. This also represents a conservative choice.

\medskip

The data adopted in our analysis are shown in Fig.~\ref{fig:fluxvsb} (adapted from Fig.~5 of \cite{Bouchet:2011fn}). They are divided into 5 energy bands (27$-$49 keV, 49$-$90 keV, 100$-$200 keV, 200$-$600 keV and 600$-$1800 keV), each featuring 21 bins in latitude (15 in the case of the fifth band). The longitude window is $-23.1^\circ < \ell < 23.1^\circ$ for the first four bands, and $-60^\circ < \ell < 60^\circ$ for the fifth one. 
\bigskip

For each DM annihilation channel, we compute the total photon flux from DM annihilation in each energy band and per each latitude/longitude bin as discussed in Sec.~\ref{sec:computations}. In Fig.~\ref{fig:fluxvsb} we show as black dots one example of the predicted photon flux which refers to a 150 MeV DM particle annihilating into $\mu^+\mu^-$. We see that the DM flux increases approaching the central latitude bins, as expected, because of the larger DM density towards the center of the Galaxy. If the emission were simply  due to FSR, the angular profile would strictly follow the DM density profile squared. This is actually not the case here, since the ICS signal, which is dominant in the configuration considered here, is molded by the spatial dependences of the $e^\pm$ energy losses and of the density of the target radiation fields. For instance, the DM flux in the very central latitude bin, which corresponds to lines of sight crossing the Galactic Plane, is significantly lower compared to the neighboring ones: this is due to the fact that the energy losses are large for electrons in the plane as a consequence of the high gas density, and therefore their ICS emissivity is suppressed. As discussed above, we remove (`mask') the 3 central latitude bins in order to exclude most of the signal from the GP~\footnote{For the first 4 energy bands, the 3 central bins cover the interval $-3.9^\circ < b < 3.9^\circ$, which corresponds to masking all lines-of-sight passing within $\sim$0.6 kpc above and below  the vertical of the Galactic Center.}. 

\medskip

We then proceed to derive constraints on Dark Matter in two different ways: we first derive conservative bounds by not including any astrophysical galactic X-ray emission; we then derive more optimistic limits by adopting a model for the astrophysical background and adding a DM component on top of it. We discuss the two strategies in turn in the following.

\medskip

{\bf Conservative constraints.} From the comparison between the predicted DM flux $\Phi_{{\rm DM}\gamma}$ and the measured flux $\phi$ we  derive constraints on the DM annihilation cross section by requiring that the former does not exceed the latter by more than an appropriate amount. More precisely, we proceed as follows. We define a test statistic 
\beq
\label{eq:chi2plus}
\chi^2_{>} = \sum_{\rm bands} \ \sum_{i \in \{{\rm b \, bins}\}} \frac{({\rm Max}[(\Phi_{{\rm DM}\gamma,i}(\langle \sigma v\rangle)-\phi_i),0])^2}{\sigma_i^2}
\eeq
where the first sum runs over the five {\sc Integral} energy bands and the second over the latitude bins (except the 3 central ones) and $\sigma_i$ is the error bar on the $i$th data point. $\chi^2_{>}$ corresponds to computing a global `effective' $\chi^2$ that includes only the data bins where the DM flux is higher than the measured value. This means that bins where the predicted DM flux is smaller that the observed one are not considered incompatible with the observations. The DM flux starts to introduce tension only when it exceeds the data points, and this tension progressively increases with the DM annihilation cross section. For each DM mass (raster scan), this is equivalent to having a number of unconstrained (non negative) nuisance parameters for the background in each bin, which makes our statistic equivalent to a $\Delta\chi^2$, distributed as a $\chi^2$ with 1 degree of freedom (see also \cite{Ackermann:2015tah} for a similar approach). We then impose a bound on $\langle \sigma v\rangle$ when $\chi^2_{>} = 4$, which corresponds to a $2\sigma$ bound.\footnote{For a detailed discussion of similar test statistics, see e.g.~Ref.~\cite{Cowan:2010js}, especially Sec.~2.5 and Sec.~3.7.}

\medskip

{\bf Optimistic constraints.} Ref.~\cite{Bouchet:2011fn} also provides templates of the different astrophysical emission processes, computed within {\sc GalProp} \footnote{\url{https://galprop.stanford.edu}},  as an attempt to explain the measured flux.Taken collectively, these astrophysical emissions constitute a background flux $\phi_B$. Including this in our analysis clearly leads to stronger constraints, since the room for exotic components is reduced. To derive DM bounds, we adopt the following (standard) procedure. 
We multiply the astrophysical background flux by an overall, energy-independent\footnote{We explored the possibility of having a different normalization of the background in the different energy bands. This assumption turned out to only mildly affect the overall constraints in the DM parameter space.} normalisation factor $N_B$. We consider the $\chi^2$ statistic 
\beq
\label{eq:chi2}
\chi^2= \sum_{\rm bands} \ \sum_{i \in \{{\rm b \, bins}\}} \frac{(\Phi_{{\rm DM}\gamma,i}(\langle \sigma v\rangle) + N_B \phi_B -\phi_i)^2}{\sigma_i^2}.
\eeq
We identify the pair $(N_{B,0},\langle \sigma v\rangle_0)$ that minimizes the statistic, corresponding to the value $\chi_0$. Then, for each DM mass, we scan the values of  $(N,\langle \sigma v\rangle)$ and we impose a constraint on $\langle \sigma v\rangle$ when $\Delta \chi^2 = \chi^2 - \chi^2_0 = 4$, for any $N_B$.

\section{Results and discussion}
\label{sec:results}

\begin{figure}[t]
\begin{center}
\includegraphics[width= 0.70 \textwidth]{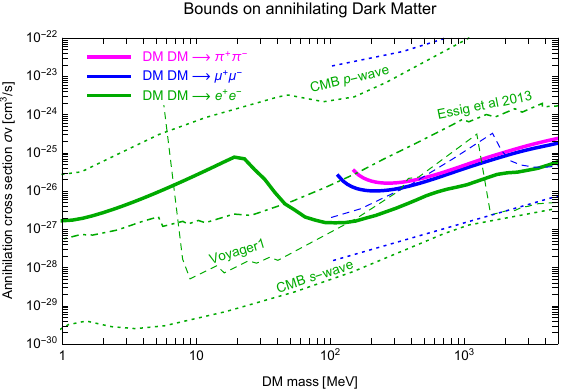}
\caption{\em \small \label{fig:bounds} Our conservative {\bfseries constraints on sub-GeV DM} from {\sc Integral} data (solid thick lines), compared to other existing bounds: from {\sc Voyager 1} $e^\pm$ data (dashed green and blue lines, from Boudaud et al.~\cite{Boudaud:2016mos}), from a compilation of X-ray data (dot-dashed green line, from Essig et al.~\cite{Essig:2013goa}) and from the CMB assuming $s$-wave (dotted green and blue lines in the lower portion of the plot, from Slatyer~\cite{Slatyer:2015jla} and Lopez-Honorez et al.~\cite{Lopez-Honorez:2013cua}) or $p$-wave annihilation (dotted green and blue lines in the upper portion of the plot, from Diamanti et al.~\cite{Diamanti:2013bia} and Liu et al.~\cite{Liu:2020wqz}; these bounds are rescaled up by a factor $(v/v_{\rm ref})^2=(220/100)^2$ since they are provided in the literature for $v_{\rm ref}=100$ km/s while we consider $v \simeq 220$ km/s in the Milky Way). For each probe, we use the color code specified in the legend: green for the DM DM $\to e^+e^-$ annihilation channel, blue for DM DM $\to \mu^+\mu^-$ and magenta for DM DM $\to \pi^+\pi^-$. When results on a channel are not present in the literature, the corresponding color is missing. For instance, there are no bounds from these probes on the $\pi^+\pi^-$ channel besides ours.}
\end{center}
\end{figure}

\begin{figure}[!t]
\begin{center}
\includegraphics[width= 0.47 \textwidth]{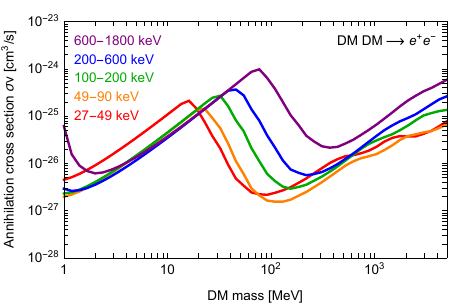} \quad
\includegraphics[width= 0.47 \textwidth]{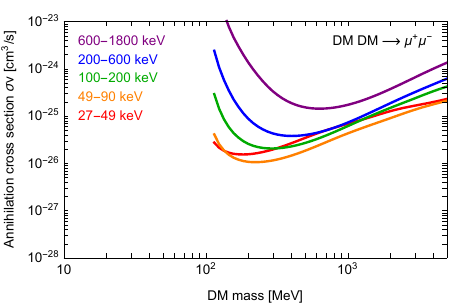} \\
\includegraphics[width= 0.47 \textwidth]{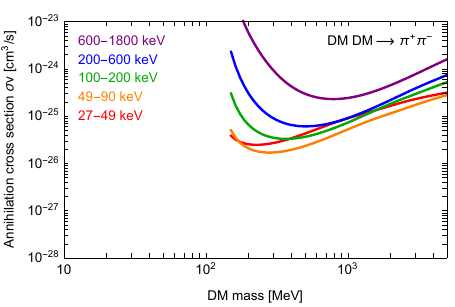} \quad 
\includegraphics[width= 0.47 \textwidth]{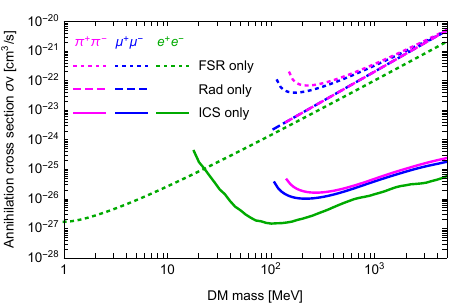} \\
\caption{\em \small \label{fig:bounds2} {\bfseries Dissection of the constraints}. First three panels: bound from each {\sc Integral} energy band, for the three annihilation channels considered. Bottom-right panel: comparison between the bounds obtained using Rad, FSR or ICS only (but combining the five energy bands).}
\end{center}
\end{figure}

\begin{figure}[!t]
\begin{center}
\includegraphics[width= 0.47 \textwidth]{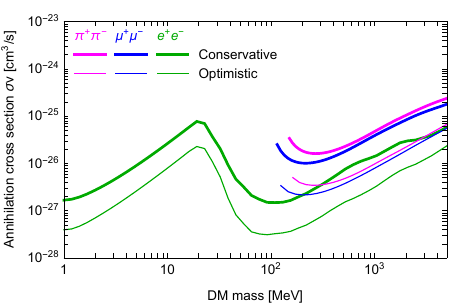} \quad 
\includegraphics[width= 0.47 \textwidth]{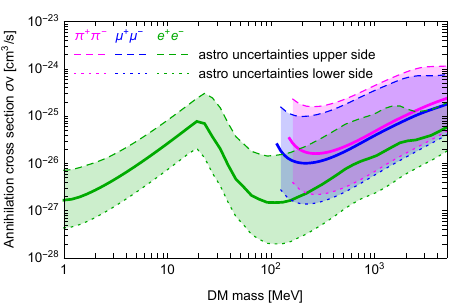} 
\caption{\em \small \label{fig:bounds3} {\bfseries Impact of the methodology and the uncertainties on the constraints}. Left: constraints with and without astrophysical background. Right: variation of the bounds due to the astrophysical uncertainties. See text for details.}
\end{center}
\end{figure}

In this section we present the constraints on Dark Matter annihilation that we derive following the procedure discussed in the previous section. Figure~\ref{fig:bounds} is our main result, as it shows the overall (conservative) bounds on the annihilation cross section $\langle \sigma v\rangle$ as a function of the DM mass for each of the three annihilation channels, obtained by combining the data from all angular and energy bins and for the total photon flux originating from both FSR and ICS processes. Figures \ref{fig:bounds2} and \ref{fig:bounds3} elucidate the impact on the bounds coming from the various elements of the analysis.

\medskip

The first three panels in Fig.~\ref{fig:bounds2} show the constraints imposed by each energy band separately. They are obtained with the same $\chi^2_{>}$ criterion defined in Eq.~(\ref{eq:chi2plus}) independently in each energy band. We can see that the dominant constraint often (although not always) comes from the 49$-$90 keV band. The highest energy band (600$-$1800 keV) almost always provides the weakest bound. The relative strength of the bounds depends on a number of factors, including statistical fluctuations in the data and the size of the errors bars in the different energy bands, but most notably from the position (in energy) of the ICS peak contribution relative to the {\sc Integral} data. Let us stress that the global constraints (in Fig.~\ref{fig:bounds}) are not just the lower envelope of the curves in these panels: they are in fact computed using all the data points in all energy bands simultaneously.

In order to understand the origin of the constraints shown in the first three panels, the last (bottom right) panel of Fig.~\ref{fig:bounds2} shows the bounds obtained by considering only the FSR, the Rad or the ICS contributions to the photon flux. As anticipated above, the ICS constraint is dominant when present. This is because for larger DM masses the ICS flux is significantly higher than the FSR and Rad one in the energy range of {\sc Integral} data (see the illustration in Fig.~\ref{fig:spectra} left). The ICS flux shifts to lower energies when the DM mass decreases, since only lower-energy $e^+e^-$ can be produced: this can be appreciated from the ICS power depicted in Fig.~\ref{fig:powerandB} (left).
As the mass becomes smaller than about 30 MeV (which can, in fact, occur only for the $e^+e^-$ channel), the ICS contribution peaks at energies lower than the band covered by {\sc Integral} and therefore becomes ineffective, thus leaving the FSR emission to dominate the constraint.

In Fig.~\ref{fig:bounds3}, the left panel shows the bounds obtained with the two procedures discussed in Sec.~\ref{sec:data}: conservative and optimistic. 
%The left-hand panel on the third row of Fig. ~\ref{fig:bounds2} displays the conservative constraints and the optimistic bounds. We find that including the astrophysical background leads to an improvement of a factor 5 in the results. However, these optimistic bounds are less conservative since the modelling of the astrophysical background remains a puzzle at the time of writing.} 
One sees that the optimistic bounds are more stringent, as expected, by about half an order of magnitude. The right panel shows the cumulative impact of the astrophysical uncertainties, as discussed in Sec.~\ref{sec:computations}. One sees that they span up to two orders of magnitude.

\bigskip

Figure~\ref{fig:bounds} also shows the comparison with other existing constraints. Essig et al.~\cite{Essig:2013goa} have derived bounds using a compilation of X-ray and soft $\gamma$-ray data from {\sc Heao-1}, {\sc Integral}, {\sc Comptel}, {\sc Egret} and {\sc Fermi}. Among the annihilation channels that we study, they consider only  $e^+e^-$. They do not include the ICS and they use {\sc Integral} data in the region $|b|<15^\circ, |\ell|<30^\circ$ rather than the latitude bins that we use (from which we exclude the Galactic plane). Their bound is comparable with ours at small masses, becoming stronger in the mass range 5-40 MeV due to the inclusion of {\sc Comptel} data, then becoming weaker for $m_{\rm DM} \gtrsim$ 50 MeV when ICS emission sets in.\footnote{Laha et al.~\cite{Laha:2020ivk}, in v1 on the arXiv, also present a result in agreement with Essig et al.~\cite{Essig:2013goa}, while in v2 the bound is no longer present.} 

Boudaud el al.~\cite{Boudaud:2016mos} have derived constraints on the $e^+e^-$ and $\mu^+\mu^-$ channel using low energy measurements by {\sc Voyager 1} of the $e^\pm$ cosmic ray flux outside of the heliosphere, using different propagation assumptions (we report the bounds of their model B, characterized by weak reacceleration). Their constraints intertwine with ours over the mass range under consideration, being stronger in the mass range 7-100 MeV and weaker otherwise.

The CMB constraints derived in~\cite{Slatyer:2015jla} are the most stringent across the whole mass range (they are given in \cite{Slatyer:2015jla} for the $e^+e^-$ channel and in the earlier study \cite{Lopez-Honorez:2013cua} for the $\mu^+\mu^-$ channel, in the mass range of interest).\footnote{Additional bounds, somehow weaker that the CMB ones, can be obtained using only the DM effect on the temperature of the intergalactic medium~\cite{Liu:2016cnk}.} However, they hold under the assumption that DM annihilation is speed-independent ($s$-wave). If the DM annihilation is instead $p$-wave, i.e.~$\langle \sigma v \rangle \propto v^2$, they weaken considerably. This can be understood qualitatively with the following argument (see \cite{Diamanti:2013bia} and the discussion in \cite{Essig:2013goa} for a more precise assessment).
The CMB constrains the energy injection from DM annihilations at high redshift (at the time of recombination or somewhat later). For $p$-wave annihilating DM, such injection was suppressed since the DM was very cold (slow) back then. In the galactic halo, at present times, DM particles move faster, as an effect of the gravitational collapse that formed large scale structures, and therefore annihilate more efficiently. In other words, a large value for the annihilation cross section at present-day is allowed as it corresponds to a much smaller value and hence a limited effect at the time of the CMB. 
Instead the bounds obtained in our analysis, and the other bounds that we report, are sensitive only to DM annihilation at the present time and therefore are independent of the $s$-wave/$p$-wave assumption if we assume, as usually done, a constant DM speed in the galactic halo. If instead we introduce a radial dependence of the DM speed, the $p$-wave bounds are affected. We have estimated that they depart from the $s$-wave ones by a factor $\mathcal{O}(40\%)$, for typical assumptions on the DM speed and density profile in the Galaxy.

{\sc Fermi} constraints as computed by the Collaboration (e.g.~\cite{Ahnen:2016qkx}) are not provided for DM masses below a few GeV, therefore we do not report them here.

\bigskip

A few other studies directly connected with our analysis have appeared in the literature in the past. Reference~\cite{McDaniel:2017ppt,Coogan:2019qpu} have introduced tools for computing signals like those in which we are interested, which however we do not employ. Reference~\cite{McDaniel:2017ppt} also derives bounds, but focuses mostly on synchrotron emission in dwarf galaxies from multi-GeV DM. Reference~\cite{Chan:2017aup} applies our approach, but considers ICS on the CMB only, focuses on $m_{\rm DM} > 1$ GeV and uses only a partial measurement from the Draco dSph galaxy. Hence there is no overlap of our results with these analyses.

\section{Conclusions}
\label{sec:conclusions}

We have derived and presented constraints on Dark Matter in the mass range 1 MeV to 5 GeV, comparing X-ray emission from the annihilation of such light DM with data from the \textsc{Integral} telescope. 
Our constraints (see Fig.~\ref{fig:bounds}) are comparable with previous results derived using X-ray data and using $e^\pm$ data from \textsc{Voyager 1}.   However, the bounds we present here are the strongest to-date on the present-day annihilation of Dark Matter for masses in the range 150 MeV to 1.5 GeV. CMB bounds remain stronger over the whole mass range, but they do rest on the assumption that the DM annihilation cross section at the time of recombination is the same as the present-day one. When this is not the case, the CMB bounds largely relax.

The strength of our constraints is due in large part to the inclusion of Inverse Compton Scattering (ICS) emission, produced by the upscattering of ambient photons by electrons and positrons produced by Dark Matter annihilation. The energy of these ICS photons is typically a few orders of magnitude lower than the DM mass, allowing us to use data from  \textsc{Integral} to help plug the `MeV gap' and produce novel constraints on sub-GeV DM. 

\medskip

In the near future, data from e{\sc Astrogam} (300 keV - 3 GeV)~\cite{DeAngelis:2017gra} or {\sc Amego} (200 keV - 10 GeV)~\cite{McEnery:2019tcm} will hopefully cover the `MeV gap' allowing us to directly probe sub-GeV annihilating DM. At the same time, upcoming full-sky data from the e{\sc Rosita} X-ray telescope (which will cover the range 0.3 - 10 keV)~\cite{2012arXiv1209.3114M} will be sensitive to the ICS emission of sub-GeV DM and will therefore likely allow us to  improve the reach of the technique we used in this paper.

\small
\subsubsection*{Acknowledgments}
M.C.~and B.J.K.~acknowledge the hospitality of the Institut d'Astrophysique de Paris ({\sc Iap}) where part of this work was done.
N.F.~and E.P.~acknowledge the hospitality of the Laboratoire de Physique Th\'eorique et Hautes Energies ({\sc Lpthe}), CNRS and Sorbonne University, where this work was started in 2016 and resumed in 2020. We thank Ennio Salvioni for comments and we acknowledge very useful discussions with John Beacom concerning In-Flight Annihilation and with Adam Coogan concerning the Radiative decays.\\

\footnotesize
\noindent Funding and research infrastructure acknowledgements: 
\begin{itemize}
\item[$\ast$] European Research Council ({\sc Erc}) under the EU Seventh Framework Programme (FP7/2007-2013)/{\sc Erc} Starting Grant (agreement n.\ 278234 --- `{\sc NewDark}' project) [work of M.C.].
\item[$\ast$] {\sc Cnrs} $80|${\sc Prime} grant (`{\sc DaMeFer}' project) [work of M.C.].
\item[$\ast$] {\sc Departments of Excellence} grant awarded by the Italian Ministry of Education, University and Research ({\sc Miur})  [work of N.F. and E.P.].
\item[$\ast$] Research grant of the Universit\`a Italo-Francese, under Bando Vinci 2020 [work of E.P.].
\item[$\ast$] Research grant {\sl The Dark Universe: A Synergic Multimessenger Approach}, No. 2017X7X85K funded by the Italian Ministry of Education, University and Research ({\sc Miur})  [work of N.F. and E.P.].
\item[$\ast$] Research grant {\sl The Anisotropic Dark Universe}, No.~CSTO161409, funded by Compagnia di Sanpaolo and University of Torino  [work of N.F. and E.P.].
\item[$\ast$] Research grant {\sc TAsP} (Theoretical Astroparticle Physics) funded by Istituto Nazionale di Fisica Nucleare ({\sc Infn})  [work of N.F. and E.P.].
\item[$\ast$] Spanish Agencia Estatal de Investigaci\'on ({\sc Aei, Miciu}) for the support to the Unidad de Excelencia Mar\'ia de Maeztu Instituto de F\'isica de Cantabria, ref. MDM-2017-0765 [work of B.J.K.].
\end{itemize}

%\newpage

\bibliographystyle{JHEP-noquotes}
\bibliography{XrayDM.bib}

\end{document}